\documentclass{article}

\usepackage{arxiv}

\usepackage[utf8]{inputenc} 
\usepackage[T1]{fontenc}    
\usepackage[colorlinks=true,allcolors=black]{hyperref}       
\usepackage{url}            
\usepackage{booktabs}       
\usepackage{amsfonts}       
\usepackage{nicefrac}       
\usepackage{microtype}      
\usepackage{lipsum}		
\usepackage{graphicx}
\usepackage[numbers]{natbib}
\usepackage{doi}
\usepackage{float} 

\title{Evaluating GPT-4 at Grading Handwritten Solutions in Math Exams}

\author{Adriana Caraeni\\
University of Massachusetts Amherst\\
acaraeni@umass.edu\\
\And
Alexander Scarlatos\\
University of Massachusetts Amherst\\
ajscarlatos@cs.umass.edu\\
\And
Andrew Lan\\
University of Massachusetts Amherst\\
andrewlan@cs.umass.edu
}

\date{}

\renewcommand{\headeright}{}
\renewcommand{\undertitle}{Published in LAK25: The 15th International Learning Analytics and Knowledge Conference}
\renewcommand{\shorttitle}{\textit{Evaluating GPT-4 at Grading Handwritten Solutions in Math Exams} }

\begin{document}
\maketitle

\begin{abstract}
Recent advances in generative artificial intelligence (AI) have shown promise in accurately grading open-ended student responses. However, few prior works have explored grading handwritten responses due to a lack of data and the challenge of combining visual and textual information. In this work, we leverage state-of-the-art multi-modal AI models, in particular GPT-4o, to automatically grade handwritten responses to college-level math exams. Using real student responses to questions in a probability theory exam, we evaluate GPT-4o's alignment with ground-truth scores from human graders using various prompting techniques. We find that while providing rubrics improves alignment, the model’s overall accuracy is still too low for real-world settings, showing there is significant room for growth in this task.
\end{abstract}

\keywords{Automated Grading \and Handwritten Math Solutions \and Large Language Models \and Multi-Modal Models}

\section{Introduction}

Automated scoring is a key challenge to enable the deployment of open-ended questions at scale. Researchers have studied the problems of automated essay scoring~\cite{attali2006automated} and automated short-answer grading~\cite{burrows2015eras} extensively, often using AI. Fine-tuning language models, such as BERT, has been effective on these tasks~\cite{zhang2022automatic}; others have recently explored prompting large language models (LLMs) for automated scoring~\cite{stahl2024exploring}. One important setting in automated scoring is student handwriting: in many practice and assessment settings, students write down their solutions to problems on paper, which are then converted to images, common in science, technology, engineering, and math (STEM) fields~\cite{baral2023auto}. This automated scoring task is challenging: compared to scoring textual essays and short answers, images contain rich semantic, visual, and even spatial information on student thought processes, which require significant textual, mathematical, and visual reasoning capabilities from AI. Recent advances in multi-modal foundation models, especially vision-language models, have significantly advanced the textual and visual reasoning abilities of AI. In this work, we perform a preliminary exploration into automated scoring of handwritten student responses to math exams using OpenAI’s GPT-4o model. We evaluate several prompts, analyze failure patterns, and find that GPT-4o's scoring ability is significantly lower than that of human graders. One recent study also evaluates AI in handwritten math grading~\cite{liu2024ai}. However, they use responses from an optional exam where students may not put in much effort, while we use real final exam responses, yielding a more realistic data source that captures student behavior in actual test-taking settings.

\section{Experimental Setup}
\label{sec:headings}

Using an IRB-approved process, we collect a dataset of real handwritten final exam responses from a single semester of a probability theory course at a university in the United States. We emailed students from the course asking permission to use their exam responses for research; we use exams from the 18 students who gave consent. We did not collect any demographic information from students. The 120-minute exam contains 5 questions, each with 3 independent sub-parts, covering the topics of probability estimates, game theory, Markov chains, Bayes nets, and parameter estimation. Each question was scored by a single grader using rubrics they wrote to assign credit for partially correct solutions. Student written exam responses contained text, mathematical formulas, and diagrams, all critical to understanding their solutions. Students in our sample scored 89.88\% on the exam on average, with many points from partial credit on incorrect answers.

We use OpenAI’s recent GPT-4o model to assign scores to student responses. We prompt the model to grade one question at a time, providing a scanned image of the corresponding page from the student’s exam and telling the model how many points each part is worth. We experiment with 3 different prompt types: i) no context \textbf{(N)}, where the model only sees the student response, ii) correct answer \textbf{(C)}, where the model sees the student response and the correct answer for each question part, and iii) correct answer and rubric \textbf{(CR)}, where the model sees the student response, the correct answer, and the rubric for each question part. We measure how well GPT-4o can score student responses, which we refer to as alignment, by comparing its predicted scores to the ground truth scores assigned by course graders. We examine scores at the question level, resulting in 18 $\times$ 5 = 90 samples, and normalize scores between 0 and 1 based on the total points per question. We then compute the mean absolute error \textbf{(MAE)}, root mean squared error \textbf{(RMSE)}, accuracy \textbf{(Acc.)}, and Pearson’s correlation coefficient \textbf{(Corr.)} between predicted and ground truth question scores. We also show the average score assigned by graders \textbf{(Score G.)} and by the model \textbf{(Score M.)}. 

\section{Results}

\begin{table} [H]
    \caption{Average alignment by prompt type. Providing the answer and rubric performs the best.}
    \centering
    \begin{tabular}{lcccccc}
        \toprule
        Prompt Type & MAE $\downarrow$ & RMSE $\downarrow$ & Acc. $\uparrow$ & Corr. $\uparrow$ & Score G. & Score M. \\
        \midrule
        N  & 0.0940 & 0.1533 & 0.4222 & 0.2776 & 0.8988 & 0.9759 \\
        C  & 0.0989 & 0.1609 & 0.4333 & 0.5502 & 0.8988 & 0.8501 \\
        CR & 0.0766 & 0.1267 & 0.4667 & 0.6174 & 0.8988 & 0.8808 \\
        \bottomrule
    \end{tabular}
    \label{tab:average_alignment}
\end{table}

To determine how relevant context in the prompt is, we show the alignment metrics and scores averaged over all students and questions partitioned by prompt type in Table~\ref{tab:average_alignment}. We observe that CR performs the best, indicating that a correct answer and rubric is necessary for GPT-4o to grade student responses accurately. We make two observations to explain this result: i) N tends to overestimate student scores since it inaccurately judges solution correctness without a reference, and ii) C tends to underestimate student scores since it rarely assigns as much partial credit as the human graders. While CR solves these two issues, we see that its predicted scores are still off by 7.66\% on average, indicating there is significant room for improvement in the handwritten grading task. 

\begin{table} [H]
    \caption{Average alignment per question using CR, varying greatly across questions.}
    \centering
    \begin{tabular}{lcccccc}
        \toprule
        Question & MAE $\downarrow$ & RMSE $\downarrow$ & Acc. $\uparrow$ & Corr. $\uparrow$ & Score G. & Score M. \\
        \midrule
        1 & 0.0833 & 0.1302 & 0.3889 & 0.4261 & 0.9028 & 0.9083 \\
        2 & 0.1235 & 0.1697 & 0.3889 & 0.4353 & 0.8580 & 0.7716 \\
        3 & 0.1011 & 0.1430 & 0.1667 & 0.5670 & 0.8167 & 0.8211 \\
        4 & 0.0389 & 0.0850 & 0.6667 & 0.3809 & 0.9694 & 0.9639 \\
        5 & 0.0361 & 0.0825 & 0.7222 & 0.8512 & 0.9472 & 0.9389 \\
        \bottomrule
    \end{tabular}
    \label{tab:alignment_per_question}
\end{table}

To determine which types of questions GPT-4o has difficulty grading, we show the alignment metrics averaged over all students with the CR prompt type for each question in Table~\ref{tab:alignment_per_question}. We observe a large difference in performance across questions, with questions 4 and 5 generally more accurate than the others. GPT-4o falls short on questions 2 and 3 primarily because students are required to justify their answers in these questions, and the model struggles to identify when these justifications are correct. In question 1, GPT-4o tends to give full marks for faulty solutions in part 2, which is on Chebyshev’s inequality. It often cannot identify the incorrect step in these solutions, possibly because they are relatively long, and this type of problem may be infrequent in the model’s training data. We also note that model accuracy is roughly correlated with student performance (Score G.), indicating the model has more trouble identifying issues with incorrect solutions than simply identifying correct solutions. 

We also perform a qualitative analysis of the model’s outputs when using the CR prompt, identify common errors, and propose solutions to explore in future work. First, the model occasionally marks clearly correct answers as incorrect or vice versa. This may be from struggling to read student handwriting, or possibly information overload from the lengthy solutions. It may be possible to ask the model if it can clearly read and understand the student solution and defer grading if it cannot. Second, the model often cannot understand if the reasoning in a student solution is correct, then incorrectly add or remove points. It may be beneficial to provide the model with a full handwritten correct solution as reference. Finally, the model sometimes misinterprets the rubrics. For example, in question 2, the rubric implies that student justifications should reference payoff matrix values; the model often removes points via this item while human graders do not. While the rubrics work for human graders, it may be helpful to write custom rubrics that are more interpretable by the model.

\section{Discussion, Conclusion, and Future Work}

In this work, we evaluate the ability of GPT-4o, a powerful multi-modal large vision-language model, to grade real handwritten student responses in college math exams. We find that providing a correct answer and rubric as reference are necessary to improve alignment with human graders, but that GPT-4o still struggles to accurately assign scores for many reasons. In particular, we find that the model struggles to comprehend student solutions, either from i) trouble reading the student’s handwriting, ii) not knowing the true correct solution steps, or iii) incorrectly interpreting the reasoning behind a student’s response. There are many avenues for future work. First, researchers should assess GPT-4o’s performance on sub-tasks, such as transcribing or reasoning over solutions. Second, researchers should investigate if fine-tuning open-source models like Llama 3.2 can improve alignment. Finally, researchers should evaluate visual grading in more domains, such as computer science or visual arts.

\bibliographystyle{plain}
\bibliography{references}

\end{document}